\documentstyle[11pt]{article}
\input{epsf}
\begin{document}

\title{\bf Thermal and Electric Conductivities of Coulomb Crystals 
           in Neutron Stars and White Dwarfs}

\author{ {\bf D. A. Baiko and D. G. Yakovlev} \\
         {\it Ioffe Physical Technical Institute, 
              Russian Academy of Sciences,} \\
         {\it Politekhnicheskaya 26, St. Petersburg, 194021 Russia} \\
         {\it Received on May 11, 1995}}

\date{${}$}
\maketitle

\def\la{\;\raise0.3ex\hbox{$<$\kern-0.75em\raise-1.1ex\hbox{$\sim$}}\;}
\def\ga{\;\raise0.3ex\hbox{$>$\kern-0.75em\raise-1.1ex\hbox{$\sim$}}\;}
\newcommand{\om}{\mbox{$\omega$}}              
\newcommand{\th}{\mbox{$\vartheta$}}           
\newcommand{\ph}{\mbox{$\varphi$}}             
\newcommand{\ep}{\mbox{$\varepsilon$}}         
\newcommand{\ka}{\mbox{$\kappa$}}              
\newcommand{\dd}{\mbox{d}}                     
\newcommand{\vp}{\mbox{\boldmath $p$}}         
\newcommand{\vk}{\mbox{\boldmath $k$}}         
\newcommand{\vq}{\mbox{\boldmath $q$}}         
\newcommand{\vv}{\mbox{\boldmath $v$}}         
\newcommand{\vect}[1]{\mbox{\boldmath $#1$}}   
\newcommand{\vF}{\mbox{$v_{\rm F}$}}           
\newcommand{\pF}{\mbox{$p_{\rm F}$}}           
\newcommand{\kF}{\mbox{$k_{\rm F}$}}           
\newcommand{\kTF}{\mbox{$k_{\rm TF}$}}         
\newcommand{\kB}{\mbox{$k_{\rm B}$}}           


\begin{center}
                        {\bf Abstract}
\end{center}

Thermal and electric conductivities are calculated
for degenerate electrons scattered by phonons in a crystal made
of atomic nuclei. The exact phonon spectrum and the
Debye--Waller factor are taken into account.
Monte Carlo calculations are performed for body-centered cubic (bcc)
crystals made of C, O, Ne, Mg, Si, S, Ca, and Fe nuclei
in the density range from $10^3$ to $10^{11}$ g cm$^{-3}$
at temperatures lower than the 
melting temperature but higher than the temperature at which the Umklapp
processes begin to be "frozen out". A simplified method of calculation
is proposed, which makes it possible to describe the results in
terms of simple analytic expressions, to extend these expressions
to any species of nucleus, and to consider face-centered cubic (fcc)
crystals. The kinetic coefficients are shown to depend tangibly on the
lattice type. The results are applicable to
studies of heat transfer and evolution of the magnetic field
in the cores of white dwarfs and in the crusts of neutron stars. The
thermal drift of the magnetic field in the crust of a neutron star is
discussed. 

\newpage

\section{INTRODUCTION}
Numerous studies have been devoted to calculations of the thermal 
and electric conductivities of electrons scattered by phonons in
dense crystalline matter of the cores of white dwarfs and the
crusts of neutron stars (see, e.g., Yakovlev and Urpin 1980; Raikh
and Yakovlev 1980; Itoh $et al.$, 1984, 1993 and references
therein). In particular, Raikh  and Yakovlev (1982) calculated
the thermal and electric conductivities by the Monte Carlo method
taking into account the exact phonon spectrum in
a Coulomb crystal of atomic nuclei, as well as the contribution of
the Umklapp processes and the normal electron-phonon scattering processes.

Later, Itoh $et al.$ (1984, 1993) performed new calculations for
the same conditions taking into consideration the Debye -- Waller factor.
This factor describes suppression of the electron-phonon scattering
for large amplitudes of ion vibrations in crystals. In their
calculations, the authors used an approximate treatment of the
Umklapp processes. Moreover, they fitted their results by
very cumbersome formulas. In addition, all the
quoted calculations were made for bcc Coulomb crystals, whereas
fcc crystals may also exist in stellar matter (Section 6).

The aim of this study is to repeat the exact ("from the first-
principles") Monte Carlo calculations of Raikh and Yakovlev (1982)
including the Debye -- Waller factor, to study the dependence 
of the kinetic coefficients on the lattice type,
and to present the results in the form suitable for practical use.
Possible applications of the results are outlined, including 
the thermal drift of the magnetic field in the crust of a
neutron star.

\section{STATEMENT OF THE PROBLEM}
We will restrict ourselves to the case of sufficiently dense
matter, $\rho \ga AZ$ g cm$^{-3}$ ($A$ and $Z$ are the mass and charge
numbers of nuclei, respectively), where the electron gas is
degenerate and almost ideal, while the atoms are completely ionized
by electron pressure. For simplicity, we consider only one
atomic species.

The state of degenerate electrons is described by the Fermi
momentum $\pF = \hbar k_{\rm F} = \hbar (3 \pi^2 n_e)^{1/3}$ 
or by the relativistic parameter
\begin{equation}
        x = {\pF \over m_e c} \approx
            1.009 \left({\rho_6  \over \mu_e} \right)^{1/3},
\label{eq:x}
\end{equation}
where $n_e$ is the electron number density, $m_e$ is the electron
mass, $\rho_6$ is the density in units of $10^6$ g cm$^{-3}$, and
$\mu_e$ is the number of baryons per electron. The electron gas is
nonrelativistic ($x \ll 1$) for $\rho \ll 10^6$ g cm$^{-3}$ and becomes
ultrarelativistic ($x \gg 1$) for $\rho \gg 10^6$ g cm$^{-3}$.

The state of ions (nuclei) is characterized by the parameter of
nonideality $\Gamma \equiv Z^2 e^2/(a \kB T)$, where 
$a = [3/(4 \pi n_i)]^{1/3}$ is the mean separation between the
nuclei, $n_i$ is the number density of the nuclei, and $\kB$ is the
Boltzmann constant. At moderately high temperatures, the nuclei
form a Coulomb crystal. A bcc crystal, which is studied in the
main part of this paper (a fcc crystal is analyzed in Section
6), is the most tightly bound crystal (see, e.g., Brush $et al.$
1966). We will consider the temperature range $T_U< T < T_m$.
Here, $T_U \sim T_p Z^{1/3} e^2/(\hbar \vF)$ 
is the temperature below which the Umklapp processes are frozen out
(see below), $\vF$ is the Fermi velocity of the electrons,
and $T_m$ and $T_p$ are the melting and ion plasma temperatures,
respectively,
\begin{eqnarray}
       T_m & = & {Z^2e^2 \over a \kB \Gamma_m} \approx
               1.323 \times 10^5 Z^{5/3} \,
               \left( {\rho_6 \over \mu_e} \right)^{1/3}
               {172 \over \Gamma_m} \; {\rm K},
      \label{eq:Tm} \\
       T_p &  = & {\hbar \omega_p \over \kB} \approx
               7.832 \times 10^6 \,
               \sqrt{{Z \rho_6 \over A \mu_e}} \; {\rm K}.
\label{eq:Tp}
\end{eqnarray}
The crystal melts at $\Gamma = \Gamma_m \approx 172$ 
(Nagara $et al.$ 1987). Under conditions of study,
$T_m$ is significantly lower than the electron degeneracy temperature.
Strictly speaking, the melting temperature of crystals made
of the lightest nuclei (H, He) decreases at high densities due to strong
zero-point vibrations of the nuclei (Mochkovitch and Hansen 1979).
We will not consider these nuclei.
The plasma temperature $T_p$ is determined by the ion plasma
frequency $\omega_p = \sqrt{4 \pi Z^2 e^2 n_i/m_i}$, where 
$m_i$ is the nuclear mass. Note that the Debye temperature of
the crystal is $T_{\rm D} \approx 0.45 T_p$ (Carr 1961). 
At $T \ga T_U$, the electron-phonon scattering can be treated in the
approximation of almost free electrons. If $Z \gg 1$, the main
contribution to the scattering comes from the Umklapp processes.
The matter is a good conductor; there are many free electrons per one ion.
The latter circumstance clearly distinguishes dense
stellar matter from "terrestrial" metals (see, e.g., Yakovlev and
Urpin 1980). At $T \ll T_U$, the character of the electron-phonon
scattering changes (Raikh and Yakovlev 1982). The Umklapp processes
are frozen out: as the temperature decreases,
these processes occur only
in those cases where the electron momenta before and after
scattering lie near intersections of the electron Fermi
surface with the boundaries of the Brillouin zones
(where the electron energy spectrum contains an energy gap).
We do not consider temperatures $T \ll T_U$ as they are usually not important
for applications.

The thermal $\kappa$ and electric $\sigma$ conductivities are
conveniently expressed in terms of the electron effective
collision frequencies $\nu_\kappa$ and $\nu_\sigma$:
\begin{eqnarray}
      \kappa & = & \frac{ \pi^2 \kB^2 T n_e}{3 m_\ast \nu_\kappa}
             \approx 4.04 \times 10^{15} x^2 \beta T_6
             \left( \frac{10^{16} {\rm s}^{-1}}{\nu_\kappa} \right)
             \; \; {\rm \frac{erg}{cm \; s \; K}},
      \nonumber \\
      \sigma & = & \frac{ e^2 n_e}{m_\ast \nu_\sigma}
             \approx 1.49 \times 10^{22} x^2 \beta
             \left( \frac{10^{16} {\rm s}^{-1}}{\nu_\sigma} \right)
             \; \; {\rm \frac{1}{s}}.
\label{eq:kappasigma}
\end{eqnarray}
Here, $\beta = \vF/c = x /\sqrt{1+x^2}$, $m_\ast = m_e \sqrt{1+x^2}$.

To calculate $\nu_\kappa$ and $\nu_\sigma$, we use the well-known
variational method (Ziman 1962) with the simplest trial
functions that describe deviation of the electron distribution
from equilibrium. For astrophysical conditions, this method was
used, for example, by Flowers and Itoh (1976) and all their
successors. At $T \ga T_p$, this method yields an exact result, while
at $T \ll T_p$, it gives an error of no more than a few percent in
calculations of $\nu_\sigma$ and no more than a few dozen percent
in calculations of $\nu_\kappa$.

We describe the electron states using the scheme of extended
Brillouin zones in the approximation of free electrons. In this
case, the Fermi surface and the dispersion relation take the same
form as those for the free electron gas. We obtain
\begin{equation}
      \nu_{\sigma,\kappa} =
          { e^2 \over \hbar \vF} \, {\kB T \over \hbar} F_{\sigma, \kappa}
          \approx 0.955 \times 10^{15}
          {T_6 \over \beta} F_{\sigma, \kappa} \; \; {\rm s}^{-1},
\label{eq:nu}
\end{equation}
where we have introduced convenient dimensionless functions
\begin{eqnarray}
      F_{\sigma, \kappa} & = & {2 \over t^2 S^2} \,
             \int_S \int_S \,
             {\dd S \, \dd S' \over q^4 |\varepsilon(q)|^2}
             \left[ 1 -  { \beta^2 q^2 \over 4 k_{\rm F}^2 } \right]
             {\rm e}^{-2W(q)}
      \nonumber \\
           & \times &
            |f(q)|^2
            \sum_s [ \vq \vect{e}_s(\vk)]^2
            {{\rm e}^{z} \over ({\rm e}^{z} -1 )^2} g_{\sigma, \kappa},
\label{eq:F}
\end{eqnarray}
in which
\begin{eqnarray}
      t & = & { T \over T_p} = 0.128 T_6
            \left( {A \mu_e \over Z \rho_6} \right)^{1/2}, \; \; \;
            z \equiv z_s = {\hbar \omega_s(\vk) \over \kB T},
     \nonumber \\
     g_\sigma & = & q^2, \; \; \; g_\kappa=
           q^2 - {q^2 z^2 \over 2 \pi^2} + {3 \kF^2 z^2 \over \pi^2},
\label{eq:t}
\end{eqnarray}
Here $T_6$ is the temperature in units of 10$^6$ K, and
$s$ = 1, 2, 3 enumerates three branches of phonon vibrations.
Integration in (\ref{eq:F}) is carried out over all possible positions of
the momenta $\vp$ and $\vp'$ of an electron before and after scattering
on the Fermi surface; $S = 4 \pi \kF^2$ is the Fermi-surface area.
The quantity $\hbar \vq = \vp - \vp'$ is the electron 
momentum transfer in a scattering event, and $\vk$ and $\vect{e}_s(\vk)$ 
are, respectively, the wave vector and the polarization unit 
vector of a phonon excited or absorbed by an electron: 
$\pm \vk = \vq - \vect{K}$, where $\vect{K}$ is such a reciprocal lattice
vector that $\vk$ lies in the first Brillouin zone. For $\vk = \vq$ 
($\vect{K}$=0), here occur the so-called normal scattering processes,
while for $\vk \neq \vq$ ($\vect{K}\neq 0$), one has the Umklapp processes.
The quantity $\varepsilon(q)$ in (\ref{eq:F}) is the static longitudinal
dielectric function of the electron gas. It describes
the screening of the Coulomb
potential of nuclei by electrons. The quantity $f(q)$ is the nuclear
form-factor that takes into account finite size of atomic nuclei. Finally,
$W(q)$ is the Debye-Waller factor (see, e.g., Davydov (1983))
\begin{equation}
        2 W(q) = 2W(\vq) = {\hbar \over 2 m_i n_i}
                \sum_s \int \, { \dd \vk \over (2 \pi)^3 }
                { [\vq \vect{e}_s(\vk)]^2 \over
                 \omega_s(\vk)}
                {\rm cth} \left( {z_s \over 2} \right),
\label{eq:DWgeneral}
\end{equation}
where integration is carried out over the first Brillouin zone.
Calculations show that under the conditions of interest, it is sufficient
to set
\begin{equation}
       2W(q) = {1 \over 3} q^2 r_0^2, \; \; \;
       r_0^2 = {3 \hbar^2 \over 2 m_i k_{\rm B} T} \,
               \left\langle  {1 \over z_s}\, {\rm cth}\,
               \left( { z_s \over 2} \right) \right\rangle ,
\label{eq:DW}
\end{equation}
where $r_0^2$ is the rms deviation of an ion at a crystal site. The
angular brackets denote averaging over phonon frequencies and
polarizations:
\begin{equation}
     \langle f_s(\vk) \rangle =
     {1 \over 3 V_{\rm B} } \sum_s \int_{V_{\rm B}} \, \dd \vk \,
     f_s(\vk),
\label{eq:averaging}
\end{equation}
Here, $V_{\rm B}$ is the volume of the Brillouin zone. At $t \gg 1$, we have
$\langle z^{-1} {\rm cth} \,(z/2) \rangle = 2 u_{-2}t^{2}$, 
while at $t \ll 1$, $\langle z^{-1} {\rm cth} \,(z/2) \rangle = u_{-1}t$.
In this case $u_n = \langle (\omega / \omega_p)^n \rangle$ are the frequency
moments of the phonon spectrum.

According to (\ref{eq:F}) and (\ref{eq:DW}), 
the Debye--Waller factor weakens the electron-phonon interaction
and increases the thermal and electric conductivities. Clearly,
it is important for strong ion vibrations in the lattice -- 
near the melting point (when thermal vibrations are especially strong) 
and at high densities of the matter (when zero-point vibrations are strong).

\section{METHOD OF CALCULATION}
Thermal and electric conductivities are calculated by 4D
integration in (\ref{eq:F}) over all orientations of the 
electron momenta $\vp$ and $\vp'$ on the Fermi surface. The
Monte Carlo method has been used for numerical integration.

Initially we have calculated the phonon spectrum for a bcc
Coulomb crystal immersed in a uniform compensating 
electron-charge background. The frequencies $\omega_s(\vk)$ and 
polarization unit vectors $\vect{e}_s(\vk)$ of phonons of three
modes ($s$ = 1, 2, 3) are determined from the system of equations
composed of the elements of the dynamic matrix (Kohen and Keffer
1955; Carr 1962). Extensive tables of the dynamic matrix elements 
have been calculated for values of $\vk$ in different
points of a primitive cell of the first Brillouin zone. 
To determine the phonon parameters for an arbitrary wave vector $\vk$,
we have interpolated the tabulated data and then solved 
the system of equations. Near the center of the Brillouin zone, 
for $k \ll q_D$ (where $q_D = (6 \pi^2 n_i)^{1/3}$ is
the radius of the sphere with a volume equal to that of the
Brillouin zone), we have used the well-known asymptotic expressions
(Kohen and Keffer 1955) for the dynamic matrix elements.
In the approximation of a uniform electron background, for $k \ll q_D$,
two phonon modes ($s$ = 1 and 2) prove to be acoustic ($\omega_s \sim 
\omega_p k/q_D$), while the third ($s$ = 3) is optical ($\omega_s 
\approx \omega_p$). Response of the electron background is 
important only for the frequency of the third mode for $k \ll q_D$.
Under the action of the response, the optical mode transforms
into an acoustic one (Pollock and Hansen 1973). 
To allow for this effect, the frequency $\omega_3(\vk)$
obtained for the uniform electron background has been
multiplied by $k/ \sqrt{k^2+\kTF^2}$, where
\begin{equation}
     \kTF = \kF \sqrt{{4e^2  \over \pi \hbar \vF}}
\label{eq:kTF}
\end{equation}
is the inverse radius of electron screening of a charge in
plasma.

In our calculations, we have used the static longitudinal 
dielectric function $\varepsilon](q)$ of a degenerate relativistic 
electron gas obtained by Jancovici (1962). 
The simplest (see, e.g., Itoh $et al.$ 1993) form-factor has been chosen,
\begin{equation}
       f(q) = \frac{3}{(q r_c)^3}
               \left[ \sin(q r_c) - q r_c \cos(q r_c) \right],
\label{eq:formfactor}
\end{equation}
corresponding to the uniform charge distribution in the proton
core (of  radius $r_c$) of an atomic nucleus. In the density range of
interest (see below), the radii of nuclei do not change under the
ambient pressure, and we can use the standard formula $r_c = 1.15 
\times 10^{-13} A^{1/3}$ cm.

Initially we have calculated the Debye -- Waller factor (\ref{eq:DW}) 
for the bcc lattice.
The results closely match those of Itoh @et al.@ (1984, 1993).
However, we have obtained a simpler fitting equation:
\begin{equation}
    2W(q)  =  {q^2 \hbar \over m_i \omega_p}
              \left(1.3995 \exp(-2.7256t)+t \,
              { 4.9801+13.00 \times 61.099 t^2 \over 1+61.099t^2} \right).
\label{eq:DWexactfit}
\end{equation}
This expression gives an error less than 1\% for any $q$.
We have calculated the integrals (\ref{eq:F})
for eight elements: $^{12}$C, $^{16}$O, $^{20}$Ne,
$^{24}$Mg, $^{28}$Si, $^{32}$S, $^{40}$Ca and $^{56}$Fe.
Unlike Itoh $et al.$ (1984, 1993), we have not considered
H and $^4$He. For hydrogen, the Umklapp processes are not allowed
at all, and the kinetic coefficients require a separate treatment. 
The helium melting temperature (Mochkovitch and Hansen 1979) 
is likely to be lower than the temperature $T_U$ at which 
the Umklapp processes are frozen out, which is
beyond the scope of our study (Section 2). We have calculated
$F_\kappa$ and $F_\sigma$F@k for the electron
relativistic parameters $x$ = 0.1, 0.3, 1, 3, 10, and 30
and for the dimensionless temperatures $t$ =
0.02, 0.04, 0.08, 0.2, 0.4, 0.8, 2, 4, 8, and 20. The selected
values of $x$ correspond to the density range from approximately 
$2 \times 10^3$ g cm$^-3$. Unlike Itoh $et al.$ (1984, 1993), we
have not considered lower densities $\rho$, since the approximation of
complete ionization does not hold at these $\rho$ (see, e.g.,
Yakovlev and Urpin 1980). For carbon, we have excluded
the value $x$ = 30, because at such a high density carbon 
is instantaneously burned in pycnonuclear
reactions (Yakovlev 1994). For all elements, we have also
excluded those $t$ at which $T > 2T_m$ (see (\ref{eq:Tm})).
Calculations of $F_\kappa$ and $F_\sigma$ for each element at fixed $x$ and
$t$ have required up to 500 000 configurations in "Monte Carlo chains"
(choices of orientations of the momenta $\vp$ and $\vp'$ on 
the Fermi surface) to achieve an acceptable accuracy of 5 - 10
In all the cases of interest, an inclusion of the nuclear 
form-factor (allowance for the non-point distribution 
of the nuclear charge) has virtually no effect on the final
result. The finite sizes of nuclei become important only at
$\rho \ga 10^{12}$ g cm$^-3$.

\section{NUMERICAL RESULTS AND THEIR ANALYTIC ANALYSIS}
For practical applications, it is useful to fit the functions
$F_\sigma$ and $F_\kappa$ by simple analytic expressions similar to those
proposed by Yakovlev and Urpin (1980) and Raikh and Yakovlev
(1982). Under the conditions of interest, the main contribution to
the electron scattering comes from the Umklapp processes. The Fermi
surface is crossed by many boundaries of the Brillouin zones, and we
may set in (\ref{eq:F})
\begin{eqnarray}
         \sum_s [ \vq \vect{e}_s(\vk)]^2
       { z^n {\rm e}^{z} \over ({\rm e}^{z} -1 )^2}
       & \approx & q^2 t^2 \pi^n G_n(t),
\label{eq:Approx} \\
   G_n(t) & = & {1 \over \pi^n t^2} \left\langle  { z^n {\rm e}^z
                \over ({\rm e}^z -1)^2} \right\rangle,
\label{eq:Gn}
\end{eqnarray}
where $n$ = 0 or 2, and the angular brackets imply averaging 
(\ref{eq:averaging}). 

The functions $G_n(t)$ were calculated by Yakovlev and Urpin
(1980), who also proposed simple fit formulas:
\begin{eqnarray}
   G_0(t) & = & {u_{-2} t \over \sqrt{t^2 + a_0  }},
\nonumber \\
   G_2(t) & = & { t \over \pi^2 (t^2+ a_2)^{3/2} }.
\label{eq:GnApprox}
\end{eqnarray}
The coefficients in (\ref{eq:GnApprox}) are chosen in such a way 
to reproduce the asymptotic expressions for $t \ll 1$ and $t \gg 1$;
equations (\ref{eq:GnApprox}) describe $G_n(t)$ with an error 
less than 10
frequency moments of the phonon spectrum (Section 2). For a
bcc lattice, we have $u_{-2}^{(bcc)}=13.00$ (see, e.g., Mochkovitch 
and Hansen 1979), $a_0^{(bcc)}=0.0174$ and $a_2^{(bcc)}=0.0118$.

Using the approximation (\ref{eq:Approx}), from (\ref{eq:F}) we obtain
\begin{eqnarray}
    F_\sigma & = & {G_0(t) \over k_F^2} \int_{q_{min}}^{2k_{\rm F}} \,
       q \, \dd q \, { | f(q)|^2 \over |\varepsilon (q)|^2}
       \left(1- {\beta^2 q^2 \over 4 k_F^2} \right) \, {\rm e}^{-2W},
\nonumber \\
   F_\kappa & = & F_\sigma +
       {G_2(t) \over k_F^2} \int_{q_{min}}^{2k_{\rm F}} \,
       q \, \dd q \, { | f(q)|^2 \over |\varepsilon (q)|^2}
       \left(1- {\beta^2 q^2 \over 4 k_F^2} \right)
       \left( {3 k_F^2 \over q^2} -{1 \over 2} \right) \, {\rm e}^{-2W}.
\label{eq:FApprox}
\end{eqnarray}
Following Yakovlev and Urpin (1980), we have taken into account
that the simplified expression (\ref{eq:Approx}) is valid only 
for the Umklapp processes. Therefore, the region of integration over
the electron momentum transfer $q$ in a scattering event is limited
by a minimum momentum $q_{min}$. We set $q_{min} = q_{\rm D} = 
(6 \pi^2 n_i)^{1/3}$, where $q_{\rm D}$ is
the radius of a sphere with the volume $V_{\rm B}$.
For $q \la q_{\rm D}$, the normal scattering processes dominate.

1D integrals (\ref{eq:FApprox}) are readily calculated numerically. 
However, they can also be approximately expressed in an analytic form.
Indeed, finite sizes of nuclei are unimportant, 
under the conditions of interest (Section 3), and we may take
$f(q)$ = 1. The dielectric function $\varepsilon(q)$ of an
electron gas differs slightly from unity for $q \ga \kTF$ (\ref{eq:kTF}). 
Only for $q \ll \kTF$ does the quantity $|\varepsilon(q)|^2$ sharply
increase: the electron screening becomes important, giving 
rise to an effective cutoff of the integrals for low $q \la \kTF$. 
Simple estimates show that $\kTF \ll q_{\rm D}$ in all the cases
of study except for those of the lowest density ($\rho \sim 
10^3$  g cm$^{-3}$). At $\rho \la ZA$  g cm$^{-3}$, the electron
screening would become extremely important, but our approach
becomes invalid. In particular, the condition of complete ionization 
is violated (Section 2). In calculating the integrals (\ref{eq:FApprox})
approximately, the electron screening may be taken into account by 
shifting the lower integration limit $q_{min}$. 
We set $q_{min}^2 \sim q_{\rm D}^2 + \kTF^2$ and $\varepsilon = 1$ 
for $q> q_{min}$. The integrals can then be taken:
\begin{eqnarray}
     F_\sigma & = & G_0(t)[2R_0(s_\sigma)- \beta^2R_1(s_\sigma)],
\label{eq:Fsigma} \\
    F_\kappa & = & G_0(t)[2R_0(s_\kappa)- \beta^2 R_1(s_\kappa)]
\nonumber \\
             & + & {1 \over 2} G_2(t)
                   [\beta^2 R_1(s_\kappa) - 3 \beta^2 R_0(s_\kappa)
                    -2R_0(s_\kappa)+3R_2(s_{1\kappa})].
\label{eq:Fkappa}
\end{eqnarray}
Here, we have introduced the functions
\begin{eqnarray}
    R_0(s) & = & {1 \over \alpha}
                 ( {\rm e}^{-\alpha s} - {\rm e}^{-\alpha}),
\nonumber \\
    R_1(s) & = & {2 \over \alpha^2}
           [{\rm e}^{-\alpha s}(1+ \alpha s) -
           {\rm e}^{-\alpha}(1+\alpha)],
\nonumber \\
    R_2(s) & = & {\rm E}(\alpha s) - E(\alpha),
\label{eq:R}
\end{eqnarray}
in which ${\rm E}(x)$ is the integral exponent, $\alpha=4r_0^2 \kF^2/3$ 
is defined by $r_0$ in the Debye--Waller factor(\ref{eq:DW}) , 
and $s= q_{min}^2/(2 \kF)^2$.  It is convenient to take somewhat
different lower limits $q_{min}$ in the functions $F_\kappa$ and $F_\sigma$ 
(see below). In our case, $q_{min} \ll \kF$ and $s \ll 1$. 
Note that E$(q)$ can be calculated conveniently using the simple fit
formula
\begin{eqnarray}
   {\rm E}(q) & = & \int_q^\infty \, { \dd y \over y} \, {\rm e}^{-y}
\nonumber \\
     & \approx & \exp \left( -{q^4 \over q^3 + 0.1397} \right)
     \left[ \ln \left(  1+ { 1 \over q} \right)
    - {0.5772 \over 1+2.2757q^2} \right] .
\label{eq:E}
\end{eqnarray}
With a mean error of about 1\%, the Debye--Waller parameter 
$\alpha$ can be fitted by 
\begin{equation}
     \alpha = \alpha_0  \;
             \left( {1 \over 2} u_{-1} {\rm e}^{-9.100t}+tu_{-2} \right),
     \; \; \;
     \alpha_0 = {4 m_e^2c^2 \over k_{\rm B} T_p m_i} x^2
        \approx 1.683 \sqrt{ { x \over AZ} },
\label{eq:alpha}
\end{equation}
where $u_{-1}$ is another frequency moment of the phonon spectrum. For a
bcc lattice, $u_{-1}^{(bcc)} = 2.800$ (Mochkovitch and Hansen 1979).

If $\alpha \ll 1$, the Debye--Waller factor is unimportant. Since
$s \ll 1$, in this case we obtain
\begin{equation}
    R_0(s) \approx R_1(s) \approx 1, \; \;
    R_2 \approx \ln \left( {1 \over s } \right).
\label{eq:RwithoutDW}
\end{equation}
These formulas correspond to the approximation used by Yakovlev
and Urpin (1980), as well as by Raikh and Yakovlev (1982). The
inclusion of the Debye--Waller factor weakens the electron
scattering, and the functions $R_0$, $R_1$ and $R_2$ become smaller
than (\ref{eq:RwithoutDW}). The functions (\ref{eq:R}) completely determine 
the dependence of electric and thermal conductivities on 
the Debye--Waller factor.

Let us compare the approximate analytic expressions (\ref{eq:Fsigma}) -- 
(\ref{eq:alpha}) for $F_\sigma$ and $F_\kappa$ with the results 
of our numerical calculations (Section 3). The values of $F_\sigma$
appear to be fitted by equation (\ref{eq:Fsigma})
with an error not larger than the error of the
computations if we set in the functions $R_0(s_\sigma)$ and $R_1(s_\sigma)$
\begin{eqnarray}
      s_\sigma & = & s_{\rm D} + s_{\rm TF},
\label{eq:s_sigma} \\
      s_{\rm D} & = & { q_{\rm D}^2 \over 4 k_{\rm F}^2}
                = \left( {1 \over 4Z } \right)^{2/3},
\label{eq:sD} \\
      s_{\rm TF} & = & { k_{\rm TF}^2 \over 4 k_{\rm F}^2}
                 = {1 \over 137 \pi \beta}.
\label{eq:sTF}
\end{eqnarray}

If one uses the same parameter @s for thermal conductivity
\begin{equation}
   s_\kappa = s_{1\kappa} = s_\sigma = s_{\rm D} + s_{\rm TF},
\label{eq:s_kappa}
\end{equation}
formulas (\ref{eq:Fkappa}) -- (\ref{eq:alpha}) fit the calculated 
values of $F_\kappa$ with a mean error of 10\% for all eight chemical 
elements and all $x$ and $t$. The approximation (\ref{eq:s_kappa})
is least accurate for light elements, primarily for carbon 
(for it, the mean error is about 15\%, and the maximum
error of $\approx$30\% is obtained at $x$ = 10 and $t$ = 0.2). The
accuracy of the $F_\kappa$ fit can be improved, on average, by
a factor of about 1.5, if we take
\begin{eqnarray}
   s_\kappa & = &  s_{\rm D} \zeta +  s_{\rm TF} \xi,
\nonumber \\
   s_{1 \kappa} & = & s_{\rm D} \zeta_1 + s_{\rm TF} \xi_1,
\nonumber \\
   \zeta & = & 0.101 + 0.0305 Z, \; \;
   \xi= 3.84 +{834 \over Z^3},
\nonumber \\
   \zeta_1 & = & {2.66 \over Z^{0.122}}-1, \; \;
   \xi_1=1.456-{257.4 \over Z^2}+{4874 \over Z^4}.
\label{eq:BestFit}
\end{eqnarray}
For applications, we recommend to use (\ref{eq:s_sigma}) and
(\ref{eq:s_kappa}) because their accuracy is quite acceptable. 
It is clear from the derivation of equations (\ref{eq:s_sigma}) and
(\ref{eq:s_kappa}) that they can also be applied to other chemical
elements. Equations (\ref{eq:BestFit}) are more accurate, 
but they hold only for $6 \leq Z \leq 26$.

The functions $F_\sigma$ and $F_\kappa$ were previously calculated
with allowance for the Debye--Waller factor by Itoh $et al.$ (1984,
1993). These authors started from approximate equations (\ref{eq:FApprox}),
whereas we have performed our numerical calculations using the general
but much more complicated equations (\ref{eq:F}).
Our calculations confirm the validity of simplified equations
(\ref{eq:FApprox}). Hence, our numerical results agree, 
within the calculational errors, with those of Itoh $et al.$
(1984, 1993). We have thus proved the validity of the approximate 
calculations of electric and thermal conductivities based on equations
(\ref{eq:Approx}) and (\ref{eq:FApprox}). Furthermore, we have carried
out an analytic analysis of equations (\ref{eq:FApprox}). 
This has enabled us to obtain the simple fit equations (\ref{eq:Fsigma}) --
(\ref{eq:s_kappa}) valid to all chemical elements considered. 
As follows from their derivation, these equations
are also valid for other elements (see above). Itoh $et al.$ 
(1984, 1993) also fitted their results by analytic expressions, but their
expressions are very cumbersome. They include 56 adjustable 
parameters for each element and do not allow one to extend the 
results directly to other elements. 

The quantities $F_\sigma$ and $F_\kappa$, calculated with and without
allowance for the Debye--Waller factor, are shown in Figs. 1 and 2.
Including the Debye--Waller factor increases the electric
and thermal conductivities of matter. The effect is especially
pronounced in the cases where the amplitude of ion vibrations in a
lattice cite is large, that is, near the melting point (where
thermal vibrations are strong) and at high densities  (where zero
vibrations are important; Fig. 2). At $T \sim T_m$ the Debye--Waller
factor increases the kinetic coefficients by a factor of 2 -- 4.

At $t \ga 1$, the electron-phonon scattering becomes quasi-elastic
(the energy transfer in a scattering event is much smaller
than $kB T$). Using (\ref{eq:FApprox}), (\ref{eq:R}) and 
(\ref{eq:alpha}), one can easily show that in
this case a single effective collision frequency can be introduced
for both electric and thermal conductivities:
\begin{eqnarray}
      \nu_0 & = & \nu_{\sigma} = \nu_{\kappa} =
          { e^2 \over \hbar \vF} \, {\kB T \over \hbar} F_0,
\nonumber \\
      F_0 & = & F_{\sigma} = F_{\kappa} = {2u_{-2} \over \alpha^2}
            \left[ \alpha \left(1 - {\rm e}^{-\alpha} \right)
            - \beta^2 \left(1 - (1+\alpha) \, {\rm e}^{-\alpha}
            \right) \right].
\label{eq:nu0}
\end{eqnarray}
We have taken into account that $s \alpha \ll 1$ in virtually
all the cases studied. Equation (\ref{eq:nu0}) is a generalization 
of the similar expression obtained by Yakovlev and Urpin (1980) without
allowance for the Debye--Waller factor.

For $t \gg 1$ and $\alpha \gg 1$, equations (\ref{eq:alpha}) and 
(\ref{eq:nu0}) yield the simple expression
\begin{equation}
    \nu_0 = {2 m_\ast e^4Z \over 3 \pi \hbar^3}.
\label{eq:nu00}
\end{equation}
In this limiting case, the collision frequency of electrons 
is temperature independent and close to a similar frequency
(see, e.g., Yakovlev and Urpin 1980) in the Coulomb liquid
($T > T_m$). Formally, the collision frequency (\ref{eq:nu00}) 
would be equal to the frequency in the liquid if the Coulomb 
logarithm were equal to $\Lambda = 0.5$.

\section{THERMAL DRIFT OF THE MAGNETIC FIELD}
The effective collision frequency (\ref{eq:nu0}) defines, 
in particular, the thermal drift of the magnetic field $\vect{B}$ 
in the neutron star crust (Urpin and Yakovlev 1980). 
Recall that the thermal drift is caused by the heat flux emerging 
from the stellar interior. For simplicity, we consider a weakly 
magnetized degenerate electron gas in which $\omega_B \tau_0 \ll 1$. 
Here, $\omega_B = eB/(m_\ast c)$ is the gyrofrequency
of electrons with energies equal to the Fermi energy, and $\tau_0=1/\nu_0$
is the effective relaxation time. Under these conditions, the
heat flux is almost unaffected by the magnetic field, and the
thermal drift is determined by the Hall component of the specific
thermal electromotive force. The thermal drift velocity $u$ is
parallel to the heat flux $Q$ and for $t \ga 1$ is equal to
\begin{eqnarray}
    u & = & {2 \eta Q \over n_e p_{\rm F} v_{\rm F}}
        \approx 72 \eta {\sqrt{1+x^2} \over x^5}
         \left( {T_e \over 10^6 \;{\rm K} }  \right)^4 \;\;
        {\rm m \over yr} ,
\label{eq:u} \\
    \eta & = & {1 \over 2} \; {
            \partial \ln(\tau_0(\varepsilon) \varepsilon^{-1})
           \over \partial \ln p}.
\label{eq:eta}
\end{eqnarray}
Here, we set $Q=\sigma T_e^4$, where $T_e$ is the effective temperature 
of the stellar surface. The numerical coefficient $\eta \sim 1$ is determined
by the logarithmic derivative with respect to the electron
momentum. Once the derivative in (\ref{eq:eta}) is taken, 
$p=p_{\rm F}$ should be set. Basically, $\eta$ can have different signs, 
depending on the mechanism of electron  scattering. 
In the Coulomb liquid ($T > T_m$), according to Urpin and Yakovlev 
(1980), $\eta > 0$, and the thermal drift is directed outward. 
In crystalline matter, from (\ref{eq:eta}) at $t > 1$ we obtain
\begin{eqnarray}
     \eta &= &{1 \over 2} - \beta^2 + {3 \alpha r_1 - 2 \alpha \beta^2 r_2
             +3 ( \beta^2 - \beta^4)r_0 \over 6 r_0 - 3 \beta^2 r_1} ,
\label{eq:eta1} \\
     r_0 & = & {1 \over \alpha} \left( 1- {\rm e}^{-\alpha} \right), \; \;
     r_1 = {2 \over \alpha^2} \left[1-(1+\alpha)
           {\rm e}^{-\alpha} \right], \; \;
\nonumber \\
    r_2 & = & {3 \over \alpha^3} \left[2 - (2+2 \alpha + \alpha^2)
           {\rm e}^{-\alpha} \right], \; \;
\label{eq:r}
\end{eqnarray}
Plots of $\eta$ against density in the neutron star envelope
with the temperature $T = 8 \times 10^7$ K are shown in Fig. 3. This
temperature corresponds to $T_e \sim 10^6$ K. The coefficient $\eta$
calculated for the crystal without allowance for the Debye--Waller
factor is indicated by long dashes. It can be obtained from (\ref{eq:eta1}) 
and (\ref{eq:r}) in the limit $\alpha \rightarrow 0$: $\eta = (2-3 \beta^2)/
(4-2 \beta^2)$ (Urpin and Yakovlev 1980). In this approximation,
$\eta$ is positive in a low-density electron gas 
(for $v_{\rm F}^2/c^2 <2/3 $) but becomes negative at
higher densities. However, Fig. 3 shows that at $T \sim T_m$ with
allowance for the Debye--Waller factor, $\eta$ remains positive and
close to a similar coefficient in the liquid phase. With increasing
density and at a fixed temperature, $t$ and $\eta$ decrease.
Formally, at a sufficiently high density, equation
(\ref{eq:eta1}) gives negative $\eta$, but this occurs only 
at $t \la 1$, when the formula becomes invalid.
In the validity range of (\ref{eq:eta1}) and for
typical values of the parameters, $\eta$ remains positive, and the
thermal drift is directed outward. The calculation of $\eta$ at $t < 1$
is very complicated and it is beyond the scope of this study.

\section{FACE-CENTERED COULOMB CRYSTALS}
The preceding astrophysical calculations of electric and thermal
conductivities have been all performed for body-centered Coulomb
crystals that are most tightly bound. In addition to bcc crystals,
fcc and simple cubic Coulomb crystals can in principle exist. The
binding energy of an ion in a Coulomb crystal of any type is $U=-\zeta 
Z^2e^2/a$ (Brush $et al.$ 1966), where $a$ is the ion sphere radius
(Section 2), and the coefficients $\zeta$ are equal 
$\zeta_{bcc}$= 0.895929, $\zeta_{fcc}$=0.895874 and
$\zeta_{sc}$=0.880059. The binding energies
for bcc and fcc crystals are very close to each other. The
difference in the ion energies (in units of thermal energy) is
$(U_{bcc}-U_{fcc})/(\kB T) \approx - 0.000055 \Gamma$,
where $\Gamma$ is the nonideality parameter (Section 2). Near
the melting point, $T \approx T_m$, we have $\Gamma \approx 172$
and $(U_{bcc}-U_{fcc})/(\kB T) \approx - 0.01$.
Consequently, in addition to bcc crystals, fcc crystals can be
be formed in many cases.

In Section 4, we have justified the validity of a simplified
calculation of thermal and electric conductivities based on
equations (\ref{eq:Approx}) and (\ref{eq:FApprox}).
Let us apply this approach to fcc crystals.
Using (\ref{eq:Gn}) and (\ref{eq:FApprox}), we can calculate the functions
$F_\sigma$ and $F_\kappa$. In this case, the type of lattice affects the
functions $G_0(t)$ and $G_2(t)$, as well as the Debye--Waller parameter
$\alpha$. Using the results of Kohen and Keffer (1955), we
have created a computer code for calculating the phonon spectrum. The
spectrum resembles the phonon spectrum of a bcc
crystal. In particular, there are two acoustic and 
one optical modes near the center of the Brillouin zone, 
but the frequencies of acoustic phonons are somewhat lower. 
As a result, the frequency moments $u_{-1}^{(fcc)}=4.03$
and $u_{-2}^{(fcc)}=28.8$ are larger than those for a bcc crystal. 
We have also calculated some other frequency moments: 
$u_1^{(fcc)}$=0.462, $u_4^{(fcc)}$=0.267. The integration
over the Brillouin zone in the expressions for $G_0(t)$, $G_2(t)$ 
and $\alpha$ has been carried out by the simplified method proposed by
Mochkovitch and Hansen (1979). The results can be fitted by the same
equations (\ref{eq:GnApprox}) and (\ref{eq:alpha}) as for the
bcc lattice but with other values of constants: 
$a_0^{(fcc)}$=0.00505, $a_2^{(fcc)}$=0.00461. The fit accuracy
turns out to be of the same order of magnitude.
All equations (\ref{eq:Fsigma}) - (\ref{eq:BestFit}) 
for $F_\kappa$  and $F_\sigma$ remain valid.

The plots of $F_\kappa$ and $F_\sigma$ versus $t$ for the bcc and fcc
lattices made of iron nuclei are given in Fig. 4. The curves are
similar but the peaks at $t \sim 1$ are much more pronounced for the
fcc lattice, which is a result of a softer spectrum of acoustic phonons. 
At higher $t$, the curves approach each other
since they are described by the common asymptote (\ref{eq:nu00}).

A simple cubic lattice is appreciably more weakly bound than a
bcc or fcc lattice. In addition, the phonon frequencies in some
crystallographic directions of a simple cubic lattice turn out to
be complex, i.e., such a lattice is unstable. This is a common and
well-known (Born 1940) property of cubic lattices composed of
particles interacting through central forces.

\section{CONCLUSION}
We have calculated the electric and thermal conductivities 
of degenerate relativistic electrons scattered by phonons 
in Coulomb bcc and fcc crystals made of atomic nuclei
taking into account the Debye -- Waller factor.
The results of these calculations are valid at
densities $10^3$ g cm$^{-3}  \la \rho \la 10^{11}$ g cm$^{-3}$. 
For bcc crystals, the calculations are performed by the Monte Carlo
method using the exact formulas. Our results confirm
the validity of the approximate approach used by Itoh $et al.$ 
(1984, 1993). An analytic analysis of the results has enabled
us to describe them by the simple fit expressions
(\ref{eq:FApprox}) -- (\ref{eq:BestFit}) and extend the
calculations to the fcc lattice. It is shown that thermal and
electric conductivities are sensitive to the lattice type.

One might expect our results to be also applicable at
sufficiently low densities, 1  g cm$^{-3} \la \rho \la 10^3$ g cm$^{-3}$, 
where the pressure ionization of atoms is incomplete. 
In this density range, the charge number $Z$ of nuclei should 
be replaced in all the above expressions with an effective 
charge number $Z_{eff} \le Z$ determined by electron
screening of the nuclear charge. However, no reliable calculations
of $Z_{eff}$ have been made as yet.

Our results are valid for degenerate cores of white dwarfs and
to outer degenerate envelopes of neutron stars. First of all, 
these results can be useful in studying the thermal evolution 
of these objects (cooling, nuclear burning of accreted matter) 
and the evolution of their magnetic field (ohmic dissipation, 
generation under thermomagnetic effects). In particular,
we have shown in Section 5 that an inclusion of
the Debye--Waller factor strongly affects the thermal
drift of the magnetic field in the crust of a neutron star.

The results can easily be generalized to higher densities, 
$10^{11}$ g cm$^{-3}$ $ \la \rho \la 10^{14}$  g cm$^{-3}$.
To do this, the functions $F_\sigma$ and $F_\kappa$
should be calculated with allowance for finite sizes of
atomic nuclei for realistic models of superdense matter of
subnuclear density. We plan to consider this in a separate work.

\begin{center}
                 {\bf ACKNOWLEDGMENTS}
\end{center}
We wish to thank M.E. Raikh and V.A. Urpin for helpful
discussions. This study was supported in part by the
Russian Basic Research Foundation (Project Code 93-02-2916), ESO
(Grant A-01- 068), and ISF (grant R6-A000). One of the authors
(D.A. Baiko) is also grateful to the mayor of St. Petersburg (Grant
661 for support of the youth's scientific work in the High
Education System) and to ISF (students' grant 555s).

\newpage
\begin{center}
                   {\bf REFERENCES}
\end{center}
\noindent
Born, M., Proc. Cambr. Phil. Soc., 1940, vol. 36, p. 160.\\
Brush, S.G., Sahlin, H.L., and Teller, E., J. Chem. Phys., vol.
45, p. 2102.\\
Carr, W.J., Phys. Rev., 1962, vol. 122, p. 1437.\\
Cohen, M.N. and Keffer, K., Phys. Rev., 1955, vol. 99, p. 1128.\\
Davydov, A.S., {\it Fizika Tverdogo Tela} (Solid State Physics),
Moscow: Nauka, 1983.\\
Itoh, N., Kohyama, Y., Matsumoto, N., and Seki, M., Astrophys. J.,
1984, vol. 285, p. 758; erratum 404, 418.\\
Flowers, E. and Itoh, N., Astrophys. J., 1976, vol. 206, p. 218.\\
Jankovici, B., Nuovo Cimento, 1962, vol. 25, p. 428.\\
Mochkovitch, R. and Hansen, J.-P., Phys. Lett., vol. A73, p. 35,
1979.\\
Nagara, H., Nagata, Y., Nakamura, T., Phys. Rev., 1987, vol. A36,
p. 1859.\\
Pollock, E.L. and Hansen, J.-P., Phys. Rev., 1973, vol. 8A, p.
3110.\\
Raikh, M.E. and Yakovlev, D.G., Astrophys. Space Sci., 1982, vol.
87, p. 193.\\
Urpin, V.A. and Yakovlev, D.G., Sov. Astron., 1980, vol. 24, p.
425.\\
Yakovlev, D. G., Acta Phys. Polonica, 1994, vol. B25, p. 401.\\
Yakovlev, D.G. and Urpin, V.A., Sov. Astron., 1980, vol. 24, p.
303.\\
Ziman, J.M., Electrons and Phonons, Oxford: Clarendon, 1960.\\

\newpage

\begin{figure}[t]
\epsfxsize=0.8\hsize
\centerline{{\epsfbox{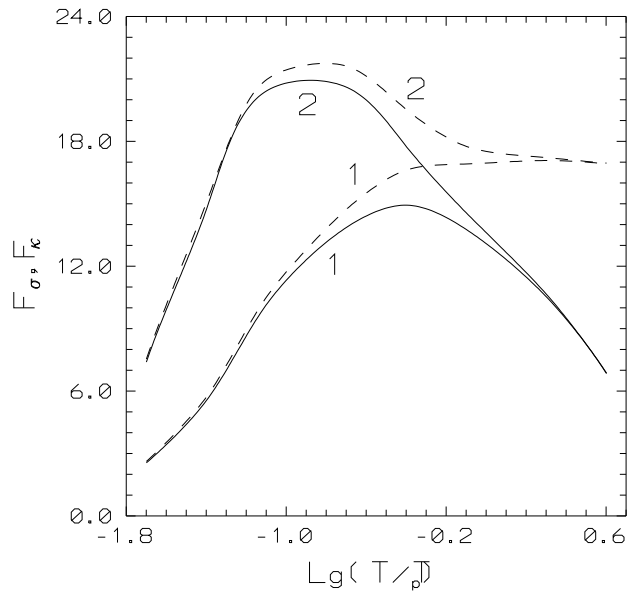}}}
\caption{\label{figure1}
$F_\sigma$ (curves 1) and $F_\kappa$ (curves 2) versus
$t = T/T_p$ for a crystal composed of $^{56}$Fe
at the density $\rho = 2.1 \times 10^6$  g cm$^{-3}$ ($x=1$) 
with (solid lines) and without (dashed lines) allowance for 
the Debye--Waller factor.
}
\end{figure}

\newpage

\begin{figure}[t]
\epsfxsize=0.8\hsize
\centerline{{\epsfbox{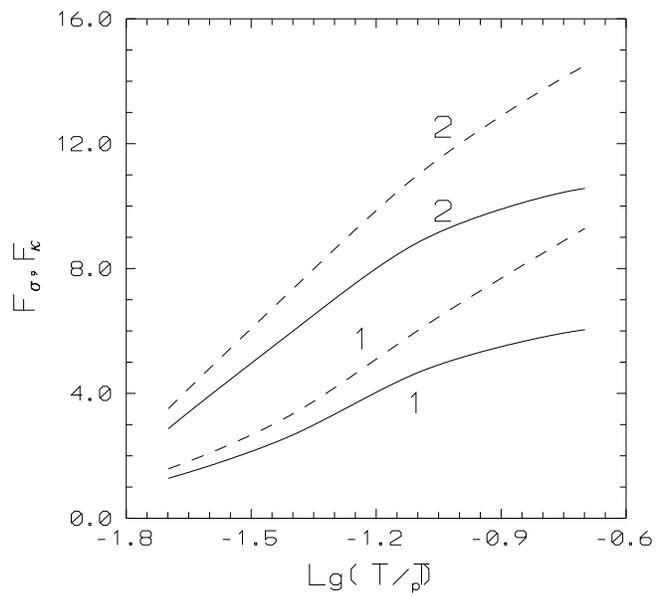}}}
\caption{\label{figure2}
Same as in Fig. 1 but for $^{12}$C nuclei at
$\rho= 5.26 \times 10^7$ g cm$^{-3}$ ($x=3$).
}
\end{figure}

\newpage

\begin{figure}[t]
\epsfxsize=0.8\hsize
\centerline{{\epsfbox{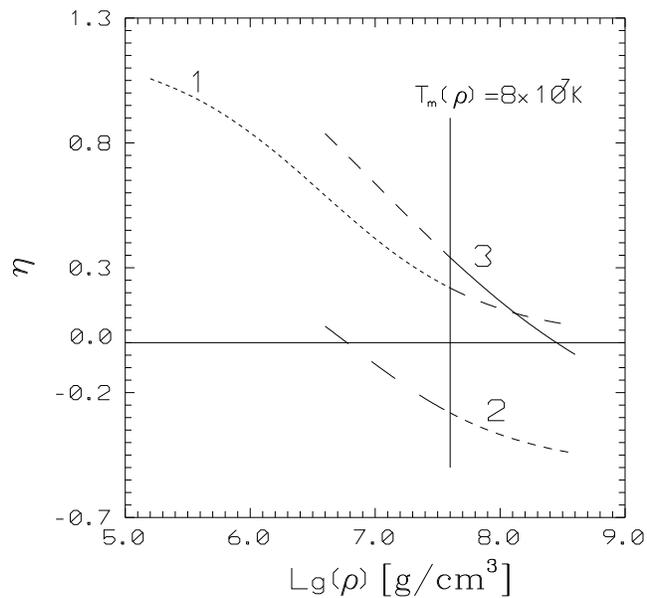}}}
\caption{\label{figure3}
The coefficient $\eta$ that determines the thermal
drift velocity of the magnetic field (see (\ref{eq:u})) as 
a function of density at $T=8 \times 10^7$ K for matter composed 
of $^{56}$Fe nuclei. Curve 1 corresponds to
the Coulomb liquid (Urpin and Yakovlev 1980), curve 2 to
a crystal without allowance for the Debye--Waller factor, and curve
3 to a crystal with allowance for this factor. The vertical line
shows the density above which solidification occurs.
}
\end{figure}

\newpage

\begin{figure}[t]
\epsfxsize=0.8\hsize
\centerline{{\epsfbox{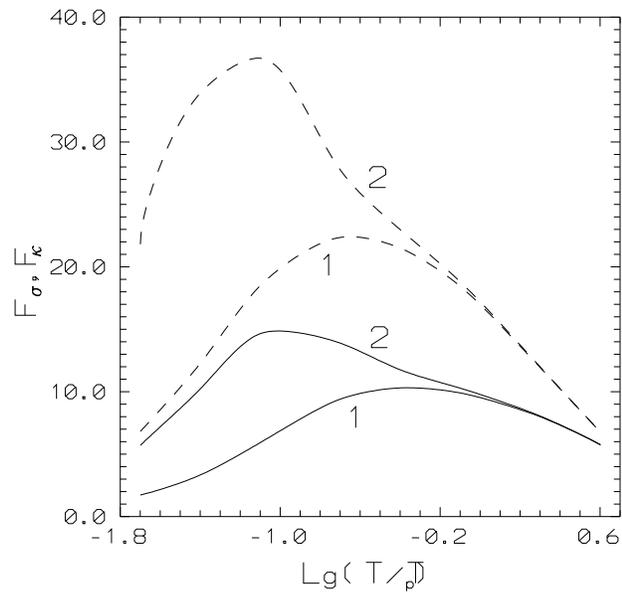}}}
\caption{\label{figure4}
Same as in Fig. 1, for bcc (solid lines) and fcc (dashed
lines) crystals with allowance for the Debye--Waller factor.
}
\end{figure}

\end{document}